\documentclass[aps,prd,preprint,tightenlines,groupedaddress,showpacs]{revtex4}
\usepackage{epsf,epsfig,graphics,graphicx}
\newcommand{\hub}{{\cal H}}

\begin{document}

\title{Inflationary dilaton-axion magnetogenesis}

\author{Shu-Lin Cheng$^1$}
\author{Wolung Lee$^1$}
\author{Kin-Wang Ng$^2$}

\affiliation{
$^1$Department of Physics, National Taiwan Normal University,
Taipei 11677, Taiwan.\\
$^2$Institute of Physics, Academia Sinica, Taipei 11529, Taiwan.}

\date{\today}

\begin{abstract}
We discuss the generation of primordial magnetic fields during inflation in the dilaton-axion electromagnetism, in which the dilaton and axion dynamics are introduced in terms of two time dependent functions of the cosmic scale factor, $I(a) F^2/4$ and $J(a) F\tilde{F}/4$, respectively, where $F$ is the electromagnetic field strength and $\tilde{F}$ is its dual. We study the form of $J(a)$ that can generate a large seed magnetic field. Although the $J(a)$ function is model dependent, the axion-photon coupling may open up a new window for a successful inflationary magnetogenesis.
\end{abstract}

\maketitle

\section{Introduction}
\label{intro}

Presently the origin still remains elusive for the observed galactic and extragalactic magnetic fields of about a few $\mu$G~\cite{kron,widrow}. It is generally believed that the so-called galactic dynamo~\cite{dynamo} mechanism must be involved to amplify a magnetic seed field $B_{\rm seed}\sim 10^{-23}$ G on a comoving scale larger than Mpc. Since the Universe is in a highly conducting plasma state through the most of its history, the ratio of the magnetic energy density and the thermal background, $\rho_B/\rho_\gamma$, remains constant along the cosmic evolution. The required $B_{\rm seed}$ is then translated into $\rho_B\simeq 10^{-34}\rho_\gamma$.

Depending on the correlation scale of the magnetic field, the seed fields can be produced inside or outside the Hubble sphere. Various astrophysical hypotheses, {\it e.g.} the Biermann battery~\cite{battery}, and cosmological scenarios, {\it e.g.} Harrison's vorticity model~\cite{harrison}, may generate large magnetic fields locally. Even if magnetohydrodynamical inverse cascades are invoked, however, the correlation scales of the resultant fields will not extend beyond 100 pc~\cite{massimorev}. Apparently these processes are incapable of explaining the observed cosmic magnetic fields which are coherent over much larger scales.

On the other hand, the correlation scale of primordial seed fields produced during inflation in the early Universe is much greater than the Hubble radius. As we know, the de Sitter like expansion is better described by the spatially flat Friedmann-Robertson-Walker (FRW) metric which conformally preserves the standard electromagnetic (EM) action~\cite{subrama}. Therefore, a viable scenario of an inflationary magnetogenesis must exploit some mechanism to break the conformal invariance of the EM field. There are several popular methods to serve the purpose. For examples, one can consider the EM field tensor $F$ coupling to the gravitational curvature scalar $R$ in the forms~\cite{tw88} of $RA^2$, $R_{\mu\nu}A^\mu A^\nu$, $RF^{\mu\nu}F_{\mu\nu}$, where $F_{\mu\nu}=\partial_\mu A_\nu - \partial_\nu A_\mu$ and $A^\mu$ is the vector potential; to scalar inflatons or dilations~\cite{tw88,binfs}; to pseudo-scalar fields like axion~\cite{carrolls};...{\it etc.} However, most of them do not produce satisfactory outcomes~\cite{gio,mandy,muk,dur10}.

Motivated by the mechanism of axial couplings~\cite{carrolls}, a scenario connecting the primordial magnetic field (PMF) to the dark energy has been developed~\cite{plb}. Assuming that an evolving pseudo scalar field is responsible for the late time acceleration of the Universe~\cite{gq}, a sufficiently large PMF can be produced inside the Hubble sphere via the spinodal instability. Furthermore, the spinodal effect is capable of generating long-wavelength fluctuations such that the created magnetic fields coherently correlate themselves over a large distance in the order of 10 Mpc.

In a previous work~\cite{pmfcjp}, we extended the axial coupling mechanism, {\it i.e.} assuming that the EM field couples to a pseudo scalar inflaton,  to explore the possibility of generating a significant PMF outside the Hubble radius . Once again we found that the spinodal instability provides a robust mechanism for generating PMF in the inflationary epoch. However, a pre-inflationary fast roll stage must be involved in order to increase the efficiency of PMF production. The correlation length of the resultant seed fields is consistent to that produced by the same mechanism inside the Hubble sphere, {\it i.e.} at a scale around 10 Mpc. Unfortunately, a further check on the energy constraint required by the inflaton field driving the de Sitter like expansion showed that such an axial coupling mechanism, when applying to the magnetogenesis process outside the Hubble sphere, would produce too much magnetic power to become a reasonable source for PMF.

In this work, we will generalize the previous work by combining both the spinodal instability in the axial coupling mechanism and the dilaton electromagnetism. This is different from the recent work that has just added EM helicity to inflationary
magnetogenesis based on dilaton electromagnetism~\cite{sorbo}. In fact, the PMF generation in our work mainly relies on the
spinodal instability.
This paper is organized as follows. In the next section, we present our formulation for the axial coupling of the EM field to a pseudo scalar field in the context of dilaton electromagnetism. We will then summarize the difficulties of inflationary magnetogenesis that makes use of dilaton and axion fields. In Sec. III, we employ the chaotic inflation scenario as a background evolution to carry out the numerical calculation and show how the PMF can be created in our model. Finally, we conclude our findings with some discussions.

\section{Dilaton-axion electromagnetism}

We consider
\begin{equation}
  \mathcal{S} =  \int d^4 x \sqrt{-g} \left[ - \frac{1}{4}F^{\mu\nu}F_{\mu\nu} - \frac{\alpha}{4 f} \chi  \, \tilde{F}^{\mu\nu} \, F_{\mu\nu} \right],
\label{emaction}
\end{equation}
where $F_{\mu \nu} = \partial_\mu A_\nu - \partial_\nu A_\mu$ is the EM field strength tensor and
$\tilde{F}^{\mu\nu} = \frac{1}{2}\epsilon^{\mu\nu\alpha\beta} F_{\alpha\beta}/{\sqrt{-g}}$ is its dual.
Note that $1/{\sqrt{-g}}$ is added to the dual tensor because $\epsilon^{\mu\nu\alpha\beta}$ is a tensor density of weight $-1$.
The pseudoscalar $\chi$ is an axion-like field, $f$ is an energy scale, and $\alpha=e^2/(4\pi)$ is the fine structure constant.
Here we assume a spatially flat FRW metric:
\begin{equation}
ds^2=-g_{\mu\nu} dx^\mu dx^\nu= a^2(\tau) (d\tau^2- d \vec{x}^2),
\label{FRW}
\end{equation}
where $a(\tau)$ is the cosmic scale factor and $\tau$ is the conformal time related to the cosmic time by $dt=a(\tau)d\tau$. The expansion rate is governed by the Hubble parameter $H \equiv \dot{a} / a= a' / a^2$, where the dot and the prime
denote derivatives with respect to $t$ and $\tau$, respectively.

After rescaling the vector potential by the coupling constant, $A_\mu \rightarrow e A_\mu$, the action~(\ref{emaction}) becomes
\begin{equation}
  \mathcal{S} =  \int d^4 x \sqrt{-g} \left[ - \frac{1}{4e^2}F^{\mu\nu}F_{\mu\nu} - \frac{1}{16\pi f} \chi  \, \tilde{F}^{\mu\nu} \, F_{\mu\nu} \right].
\label{emaction2}
\end{equation}
Now we consider a time-dependent coupling constant described by a time funciton $I(\tau)=I(\sigma(\tau))=1/e^2$ which is driven by a time-evolving dilaton field $\sigma$. In the axion sector, we assume that the axion mean field is $\chi=\chi(\tau)$. In addition, inspired by extra-dimension theories, we assume that the energy scale is related to the reduced Planck mass by $f=M_p/S(\tau)$. For example, this relation has a simple geometrical origin in many string theory constructions with the factor $S$ determined by a combination of the size of the compactification manifold, the string length, and the string coupling~\cite{axiverse}. This factor $S$ may be time dependent due to a possible temporal evolution of extra dimensions and string parameters. We define a time function $J(\tau)=S\chi/(4\pi M_p)$ to better monitor the axial effect on the processes of magnetogenesis. Subsequently, in terms of the pair of coupling functions $I(\tau)$ and $J(\tau)$, the action can be recast in the form of
\begin{equation}
  \mathcal{S} =  \int d^4 x \sqrt{-g} \left[ - \frac{1}{4}I(\tau)\,F^{\mu\nu}F_{\mu\nu} - \frac{1}{4}J(\tau)\, \tilde{F}^{\mu\nu} \, F_{\mu\nu} \right].
\label{IJaction}
\end{equation}

To derive the EM wave equation, we chose the temporal gauge, i.e. $A_{\mu} = (0, \vec{A})$, and the Coulomb gauge, i.e. $\vec{\nabla} \cdot \vec{A} = 0$. Under these gauges and the metric~(\ref{FRW}), the physical electric and magnetic fields are respectively given by
\begin{equation}
\vec{E} = -{1\over a^2} \frac{\partial \vec{A}}{\partial \tau},\quad \vec{B} = {1\over a^2}\vec{\nabla} \times \vec{A}.
\end{equation}
From the action~(\ref{IJaction}), we obtain the wave equation for the vector potential,
\begin{equation}
\frac{\partial^2 \vec{A}}{\partial \tau^2} - \vec{\nabla}^2 \vec{A}+\frac{I'}{I}\frac{\partial\vec{A}}{\partial \tau} -  \frac{J'}{I} \vec{\nabla} \times \vec{A} =0.
\label{wave}
\end{equation}
To proceed, we decompose the gauge field $\vec{A}(\tau,{\vec x})$ as
\begin{equation}
 \vec{A}(\tau,{\vec x}) = \sum_{\lambda=\pm} \int \frac{d^3k}{(2\pi)^{3/2}} \left[ \vec{\epsilon}_\lambda({\vec k}) a_{\lambda}({\vec k}) A_\lambda(\tau,{\vec k}) e^{i {\vec k}\cdot {\vec x}} + \mathrm{h.c.}   \right],
\label{decomposition}
\end{equation}
where the annihilation and creation operators obey
\begin{equation}
\left[a_{\lambda}({\vec k}), a_{\lambda'}^{\dagger}({\vec k'})\right] = \delta_{\lambda\lambda'}\delta ({\vec k}-{\vec k}'),
\end{equation}
in which $\vec{\epsilon}_\lambda$ are normalized circular polarization vectors satisfying
$\vec{k}\cdot \vec{\epsilon}_{\pm} \left( \vec{k} \right) = 0$,
$\vec{k} \times \vec{\epsilon}_{\pm} \left( \vec{k} \right) = \mp i k \vec{\epsilon}_{\pm} \left( \vec{k} \right)$,
$\vec{\epsilon}_\pm \left( -\vec{k} \right) = \vec{\epsilon}_\pm \left( \vec{k} \right)^*$, and
$\vec{\epsilon}_\lambda \left( \vec{k} \right)^* \cdot \vec{\epsilon}_{\lambda'} \left( \vec{k} \right) = \delta_{\lambda \lambda'}$.
Inserting the decomposition~(\ref{decomposition}) into Eq.~(\ref{wave}), we obtain the equation of motion for the mode functions,
\begin{equation}
  \left[ \frac{\partial^2}{\partial\tau^2} +\frac{I'}{I}\frac{\partial}{\partial\tau} + k^2 \mp k\frac{J'}{I}\right] A_{\pm}(\tau,k) = 0.
\label{f_V}
\end{equation}
Then, the energy density of the produced EM fields is governed by the vacuum expectation value,
\begin{equation}
\rho_{\rm EM}\equiv \frac{I}{2}\langle \vec{E}^2+\vec{B}^2 \rangle = \frac{I}{4 \pi^2 a^4} \int d k \,  k^2 \sum_{\lambda=\pm}\left( \vert A_\lambda' \vert^2 + k^2 \vert A_\lambda \vert^2 \right),
\label{EMenergy}
\end{equation}
where $ A_\pm'$ terms characterize the electric and the magnetic energy density of the PMF as
\begin{equation}
\rho_{B}= \frac{I}{4 \pi^2 a^4} \int d k \,  k^4 \left(  \vert A_+ \vert^2 + \vert A_- \vert^2 \right).
\label{Benergy}
\end{equation}

\section{Difficulties in inflationary magnetogenesis}
\label{diff}

As we have mentioned in Sec.~\ref{intro}, there have been many attempts to generate large-scale PMF in inflation by
invoking a breaking of the conformal invariance of the EM field.
The main hurdle that one has to overcome can be easily understood by a simple scaling argument as follows.
We need to produce PMF with magnetic energy density at the end of inflation
of order $\rho_B\simeq 10^{-34}\rho_\gamma$. Not only the produced magnetic field strength has to be big enough, but also the wavelength of the magnetic fields is of large scales. As such, EM $k$-modes are thought to be amplified and leave the horizon
at about $60$ e-foldings before inflation ends. Because of the $a^4$ dilution factor in $\rho_B$ (see Eq.~(\ref{Benergy})),
the PMF energy density is about $e^{240}$ times larger at the time of production than that at the end of inflation.
If the reheating process is rapid and efficient, the energy density of the thermal background $\rho_\gamma$ will amount to that stored in the inflaton field $\rho_\phi$ during inflation, i.e. $\rho_\gamma\simeq \rho_\phi$ right after inflation.  Therefore, at the time PMF are produced, $\rho_B\simeq 10^{-34} e^{240}\rho_\phi$, which exceeds the inflaton energy by $10^{70}$ times. This huge discrepancy already poses a serious challenge to the inflationary magnetogenesis.

Let us go back to the action~(\ref{IJaction}). When $I(\tau)=1$ and $J(\tau)=0$, it resumes Maxwellian electromagnetism,
in which EM mode functions
$ A_\pm$ in Eq.~(\ref{f_V}) have simply plane-wave solutions, as dictated by the conformal invariance of the action. This invariance prohibits the growth of EM fields in inflation. In the following, we will briefly review the efficiency of each mechanism to break
the conformal invariance of the EM field for the generation of PMF in inflation.

When $I=I(\tau)$ and $J(\tau)=0$, it is the so-called dilaton electromagnetism. In the strong EM coupling case, i.e. $ I(\tau)<1$, strong enough PMF can be produced when the effective coupling constant, which is inversely proportional to I, is extremely large
in the beginning of inflation and becomes of the order of one at the end of inflation. But one has to bear with an extremely strongly coupled EM theory that should not be trusted at all. In the weak EM coupling case, i.e. $I(\tau)>1$, no sizable PMF can be produced~\cite{muk}.

When $I(\tau)=1$ and $J(\tau)=\alpha\phi/f$, where the axion is identified as the inflaton $\phi$,  the mode equation~(\ref{f_V}) becomes
\begin{equation}
  \left[ \frac{\partial^2}{\partial\tau^2} + k^2 \mp 2aH k\xi\right] A_{\pm}(\tau,k) = 0, \hspace{5mm} \xi \equiv \frac{\alpha \dot{\phi}}{2 f H}\,.
\label{axionXi}
\end{equation}
It is well known that either one of the two modes exhibits a spinoidal instability as long as the modes satisfy
the condition, $k/(aH)<2\vert\xi\vert$, where $\xi$ is nearly a constant for a slow-roll inflation. When the inflaton rolls down the potential, these unstable modes grow exponentially by consuming the inflaton kinetic energy. It has been found that the generation of helical magnetic fields during the single-field slow-roll inflation due to this axial coupling of EM field to the inflaton leads to a blue spectrum of magnetic fields. Although the helical magnetic fields can further undergo,
during the radiation and matter dominated eras, a process of inverse cascade that transfers spectral power from small to large scales, the magnetic fields generated by such an axial coupling would not lead to a strong enough field strength on cosmological scales~\cite{dur10}. In Ref.~\cite{pmfcjp}, a pre-inflationary fast roll stage is involved in order to enhance the efficiency of helical magnetic field production. Unfortunately, when applying to the magnetogenesis process outside the Hubble sphere, it would produce too much $\rho_{\rm EM}$ that imposes too strong a backreaction on $\rho_\phi$ to
become a reasonable source for PMF.

Recently, the authors in Ref.~\cite{sorbo} have discussed the weak EM coupling case in which a new parity violating term is introduced, given by the $J(\tau)$ term with $J(\tau)\propto I(\tau)$. This parity violating term allows more freedom in tuning the amplitude of the produced helical magnetic field at the end of inflation. However, the model still relies on the inverse cascade processes to amplify a helical magnetic field at large scales during the radiation dominated epoch. As a consequence, a magnetic seed field $B_{\rm seed}\sim 10^{-18}-10^{-15}$ G can be produced if inflation occurs at an energy scale ranging from $10^5$ to $10^{10}$ GeV.

Another difficulty is that the electric field density would be larger than the total energy density of the Universe if a sufficient amount of PMF is produced. Suppose the helicity $+$-mode grows efficiently as $A_+ \propto a^\omega$, where $\omega$ is treated approximately time independent and $\omega>1$. Then, from Eqs.~(\ref{EMenergy}) and (\ref{Benergy}) we have the ratio of the electric and magnetic spectral energy densities as
\begin{equation}
\frac{\rho_E(k)}{\rho_B(k)} \sim \left(\frac{A'_+}{k A_+}\right)^2 \sim \frac{\omega^2 a^2 H^2}{k^2}.
\end{equation}
The $k$ values for large-scale PMF are about $H<k<10^4H$, so this ratio is as small as $10^{-8}\omega^2 a^2$. Going back to the end of the magnetogenesis phase that is assumed to occur at $a=e^{N_e}$, the energy densities of the PMF and the electric fields are given by, respectively,
\begin{equation}
\frac{\rho_B}{\rho_\phi} \sim 10^{-34} e^{(240-4N_e)},\quad \frac{\rho_E}{\rho_B} > 10^{-8}\omega^2 e^{2N_e}.
\end{equation}
Hence, at the end of the magnetogenesis phase we have
\begin{equation}
\frac{\rho_E}{\rho_\phi} \sim \omega^2 10^{(62-2N_e\log e)},
\end{equation}
showing that the electric energy density is larger than the total energy density of the Universe, which contradicts the conservation of energy.

\section{Dilaton-axion magnetogenesis}

In the present work, we assume an inflationary background driven by the inflaton field $\phi$.
We employ the chaotic inflation scenario~\cite{chaotic} to unravel the inflationary magnetogenesis by full numerical calculations. In such a universe, the standard slow-roll inflationary background is driven by the potential
\begin{equation}
V(\phi)={m^2\over 2}\phi^2,
\end{equation}
where the parameter $m$ characterizes the mass of the inflaton field. On the other hand, in order to solve for the ratio of the magnetic energy density to the energy density of the thermal background $\rho_\gamma$ at the end of inflation, we mimic the evolution of the thermal background by tracing out changes in the energy density of the radiation component $\rho_\gamma$. Accordingly, the background dynamics are governed by
\begin{eqnarray}
\ddot\varphi+\left(3\hub+\gamma\right)\dot\varphi+\beta^2\varphi &=& 0, \label{varphieq}\\
\dot\rho_\gamma+4\hub\rho_\gamma&=&\gamma\dot\varphi^2,
\end{eqnarray}
where $\beta^2=m^2/M_p^2$, and all other dynamical variables have been properly rescaled by the reduced Planck mass, {\it e.g.}
$\varphi=\phi/M_p$, $\hub=H/M_p$, $\dots$, {\it etc.} In Eq.~(\ref{varphieq}) the interaction of the inflaton with other light fields is modeled by a damping term $\gamma\dot\varphi$, which is not important in the slow-roll regime where $\hub$ is dominant. The Hubble parameter $\hub$ that measures the expansion rate of the universe is characterized by
\begin{equation}
\hub^2={1\over 3}\left({1\over 2}\dot\varphi^2+{\beta^2\over 2}\varphi^2+\rho_\gamma\right).
\end{equation}
Since the accelerated expansion would dilute everything that may have existed prior to the inflation, the initial value for $\rho_\gamma$ does not matter to our calculations. The initial value of the Hubble parameter at $\tau=\tau_i$ is determined by values of $\varphi_i$ and $\dot\varphi_i$ through $6\hub_i^2=\dot\varphi_i^2+\beta^2\varphi_i^2$. We take $m=1.8\times 10^{13}$ GeV for the inflaton mass, $\gamma=0.01\beta$, and $\varphi_i=-15.35$, $\dot\varphi_i=6\times 10^{-6}$ respectively for the initial position and the initial velocity of the inflaton $\varphi$.
The e-folding since the beginning of inflation is defined by $N(t)=\int_0^t H(t') dt'$ and we find that our inflation ends at $N\simeq 60$.
Since we control the production of PMF during inflation such that $\rho_{\rm EM}\ll \rho_\phi+\rho_\gamma$, the backreaction of EM field production on the inflationary background can be safely neglected.

To make our scenario of inflationary magnetogenesis more general, we do not specify any model for the dynamics of dilaton or axion during inflation. Instead, we assume that
\begin{equation}
I(\tau)=\left[\frac{a(\tau_f)}{a(\tau)}\right]^n,\quad\quad J'(\tau)=c\,H(\tau) a(\tau)^p\,,
\label{IJform}
\end{equation}
where $n$, $p$, and $c$ are constant parameters within time intervals, and $\tau_f$ denotes the time at the end of inflation. Moreover, we presuppose that $I(\tau)=1$ and $J'(\tau)=0$ for $\tau\ge \tau_f$, so the EM theory becomes Maxwellian electromagnetism after inflation. 

Now we insert $I(\tau)$ and $J'(\tau)$ in Eq.~(\ref{IJform}) into the mode equation~(\ref{f_V}). To solve for the mode functions, we set the initial conditions at $\tau=\tau_i$ as
\begin{equation}
A_{\pm}(\tau_i,k)=\frac{1}{\sqrt{I(\tau_i)}}\frac{1}{\sqrt{2k}}\,,\quad\quad A_{\pm}'(\tau_i,k)=\frac{1}{\sqrt{I(\tau_i)}}\frac{-ik}{\sqrt{2k}}.
\end{equation}
Then, we calculate the spectral magnetic energy density,
\begin{equation}
\frac{d\rho_{B}}{d\ln k}= \frac{k^5 I}{4 \pi^2 a^4}\left(  \vert A_+ \vert^2 + \vert A_- \vert^2 \right),
\label{dBenergy}
\end{equation}
and its ratio to the energy density of the thermal background at the end of inflation,
\begin{equation}
r\equiv\left[\rho_\gamma^{-1}{d\rho_B\over{d\ln k}}\right]_{\tau=\tau_f}.
\end{equation}

Here we are not going to extract that of the parameter space in our model for a successful inflationary magnetogenesis. Rather, we will give some simple examples to illuminate how and why axion dynamics can greatly facilitate the PMF production during inflation. The mode equation~(\ref{f_V}) then becomes
\begin{equation}
{\ddot A}_{\pm} + \left[(1-n)H +\ln\left({a_f}\over{a}\right){\dot n}\right]{\dot A}_{\pm}+\left[\frac{k^2}{a^2} \mp \frac{ckH a^{n+p-2}}{a(\tau_f)^n}\right]A_{\pm}=0.
\label{f2_V}
\end{equation}

As the first example,  we just take $n=1$ to efface the friction term.  As long as $k/H< ca^{1+p}/a(\tau_f)$, the mode function $A_+$ becomes unstable and grows exponentially with time.  The mode growth rate can be estimated by $A_+\propto a^\omega$, where $\omega$ is the growth index given by $\omega=[(k/H) ca^{p-1}/a(\tau_f)]^{1/2}$. During this weak EM coupling regime, the electric energy density dominates over the magnetic energy density. Specifying $c=1.0$ and $p=0.41$ leads both the electric and magnetic energy densities to grow exponentially near $N(t)\sim 45$. To avoid over-production, we have imposed a strong EM coupling regime with $n=-5.7$ after $N(t)\sim 45$ and introduced a wavenumber cutoff by assuming that $c=0$ for $k/H(\tau_i)> 300$. The strong EM coupling damps out the electric fields while freezing the magnetic fields. The wavenumber cutoff signifies the effect of backreactions due to the rapid production of high $k$-modes. Here we simply ignore these modes. It would be interesting to take into account the backreaction in the mode equation to study how the production of high $k$-modes can be self-regulated by the backreaction.

\begin{figure}[htp]
\centering
\includegraphics[width=0.8\textwidth]{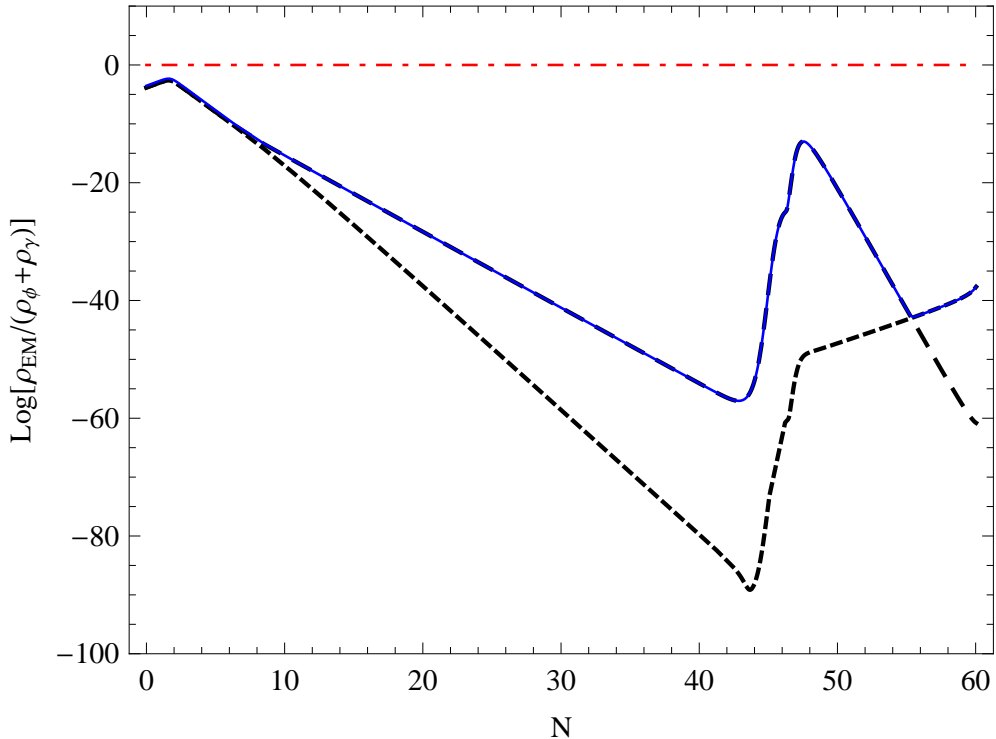}
\caption{Solid line shows the time evolution of the electromagnetic energy density over the total energy density during inflation, $\rho_{\rm EM}/( \rho_\phi+\rho_\gamma)$. The time is counted with e-foldings $N(t)$ from the beginning of inflation. The dashed (dotted) line represents the electric (magnetic) component. The horizontal dotted-dashed line signifies the conservation of energy.}
\label{fig1}
\end{figure}

\begin{figure}[htp]
\centering
\includegraphics[width=0.8\textwidth]{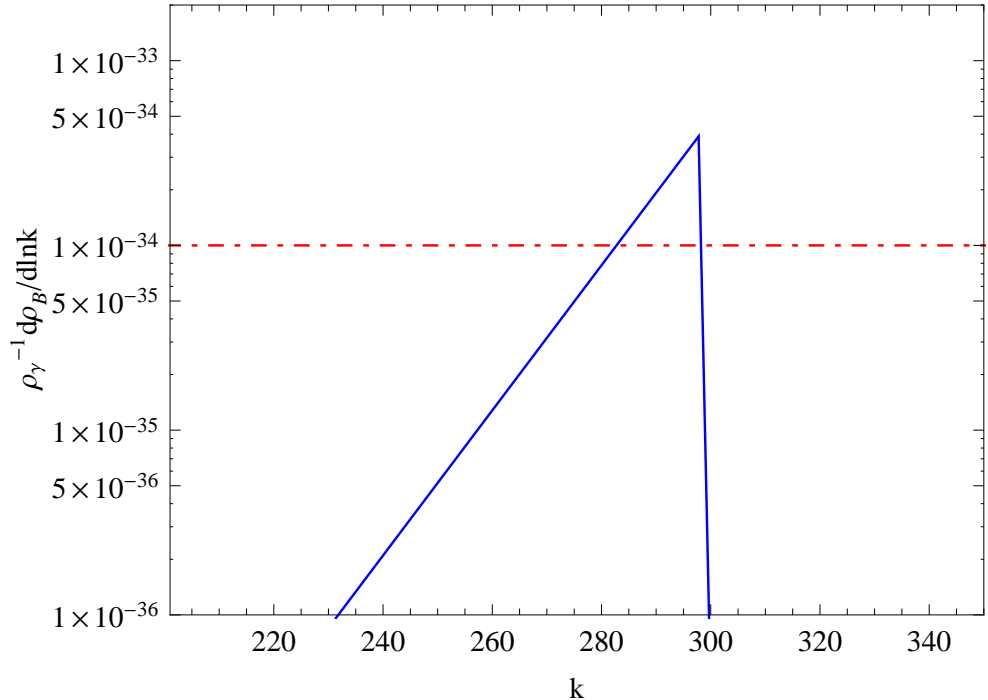}
\caption{Ratio of the spectral energy density of primordial magnetic fields to the thermal background at the end of inflation versus the wave numbers $q\equiv k/H(\tau_i)$ . The mode with $q=1$ corresponds to the size of the present Universe while the spectrum peaks at $q=300$ on a comoving scale larger than Mpc.}
\label{fig2}
\end{figure}

Figure~\ref{fig1} shows the time evolution of the electric and magnetic energy densities with $N(t)$. Also shown is the evolution of $\rho_{\rm EM}/( \rho_\phi+\rho_\gamma)$ that monitors the condition of a small backreaction. In Fig.~\ref{fig2}, the ratio $r$ is plotted against $k/H(\tau_i)$, where the length scale is set up such that $k/H(\tau_i)=1$ corresponds to the mode that leaves the horizon about $60$ e-foldings by the end of inflation, i.e. its wavelength is comparable to the size of the present Universe. The spectral magnetic energy density of the PMF shows a peak around $k/H(\tau_i)=300$, which corresponds to a comoving scale of order Mpc, because of the wavenumber cutoff. Integrating this spectrum amounts to $\rho_B\simeq 10^{-34}\rho_\gamma$ at the end of inflation. If the backreaction is included in the PMF production,  the production of high $k$-modes should be effectively reduced to a level comparable to the peak height of the spectrum.

Let us return to the axion inflation discussed in Sec.~\ref{diff}, where $I(\tau)=1$ and $J(\tau)=\alpha\phi/f$, and the growth condition is  $k/(aH)<2\vert\xi\vert$ with $\xi$ a nearly constant parameter. The axion inflation indeed corresponds to the case with $n=0$ and $p=1$, or to the case with $n=1$ and $p=0$ as far as the mode growing behavior is concerned. As a consequence, we see that the mode growth index $\omega$ with $p=0$ is suppressed by the scale factor $a$ for the modes of cosmologically interesting scales, say $k/H\lesssim 10^3$, towards the end of inflation. Now it is transparent that we need at least $p>0$ to boost the generation of PMF. Figure~\ref{fig1} shows that the $J(\tau)$ term with $p=0.41$ allows the modes of comologically interesting scales to grow sufficiently fast at a much later time ($N\sim 45$), thus circumventing the difficulty for the inflationary magnetogenesis, which has assumed that the comologically interesting modes get amplified and leave the horizon at about $N\sim 1-5$, as discussed in Sec.~\ref{diff}.

\begin{figure}[htp]
\centering
\includegraphics[width=0.8\textwidth]{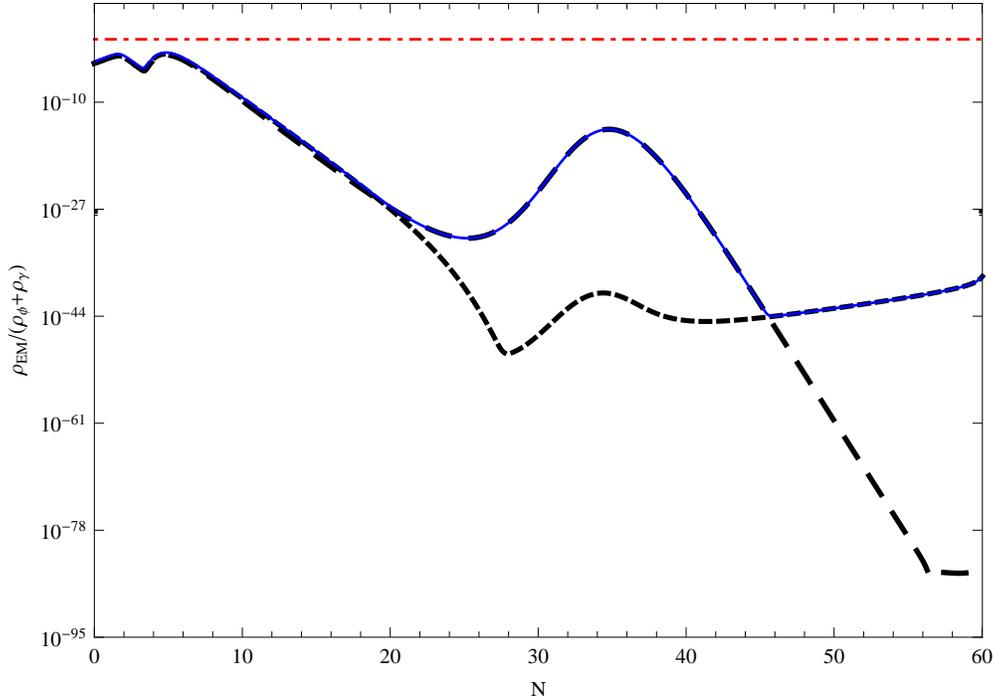}
\caption{As for Fig.~\ref{fig1} but in the strong EM coupling regime.}
\label{fig3}
\end{figure}

\begin{figure}[htp]
\centering
\includegraphics[width=0.8\textwidth]{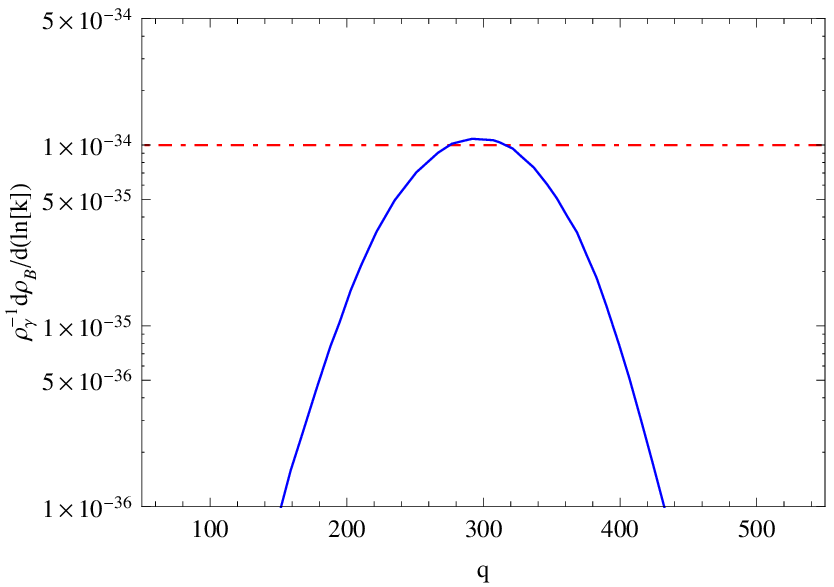}
\caption{As for Fig.~\ref{fig2} but in the strong EM coupling regime.}
\label{fig4}
\end{figure}

\begin{figure}[htp]
\centering
\includegraphics[width=0.8\textwidth]{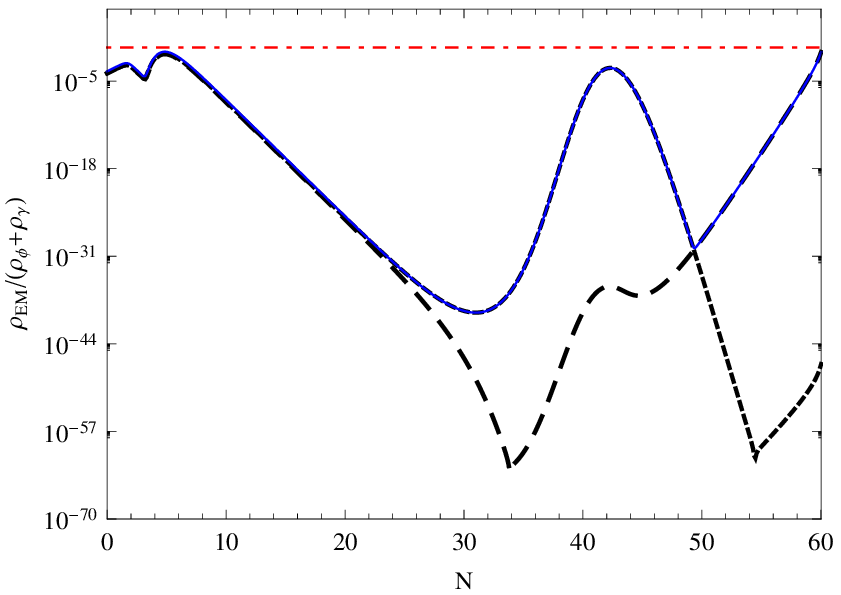}
\caption{As for Fig.~\ref{fig1} but in the weak EM coupling regime.}
\label{fig5}
\end{figure}

\begin{figure}[htp]
\centering
\includegraphics[width=0.8\textwidth]{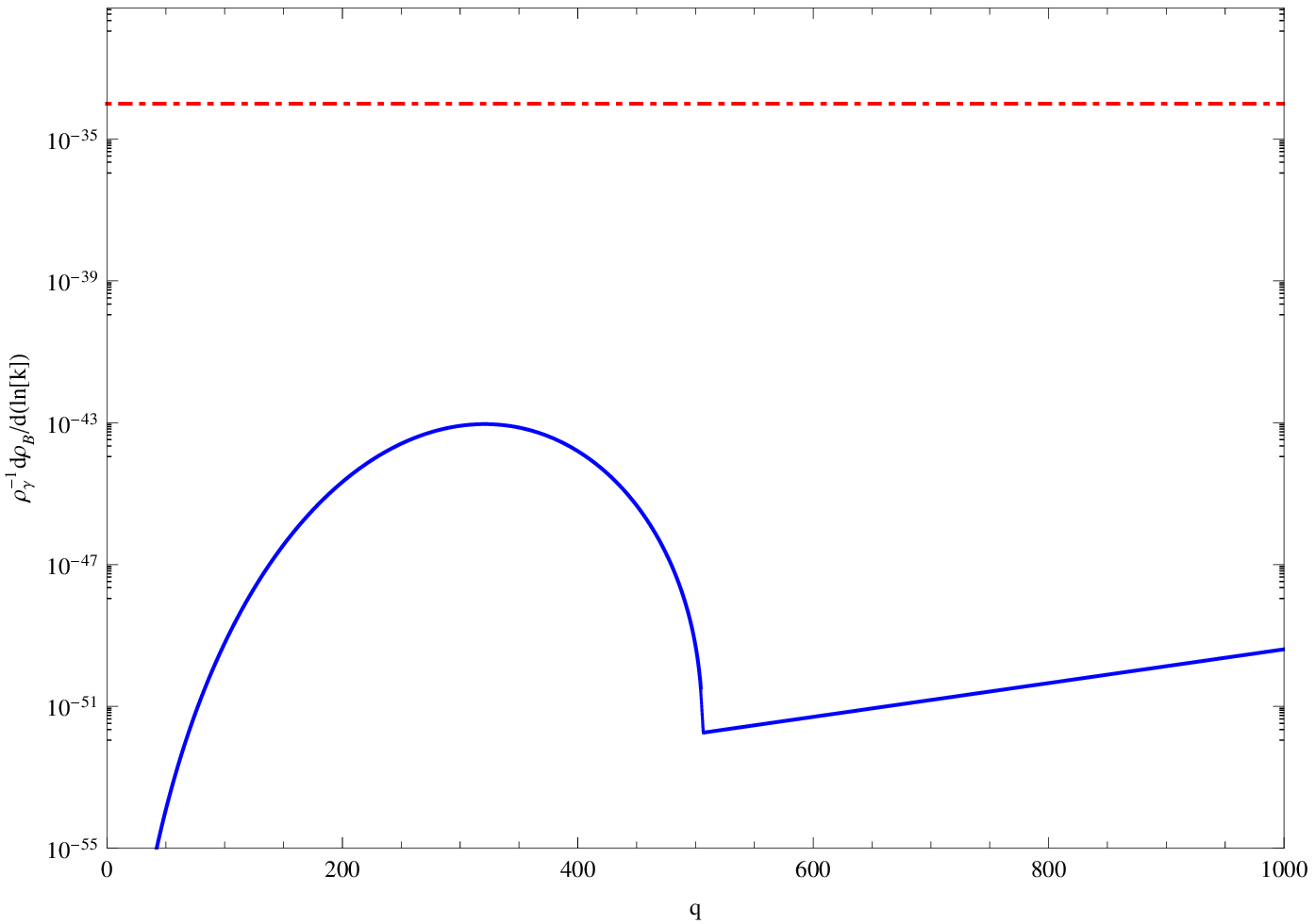}
\caption{As for Fig.~\ref{fig2} but in the weak EM coupling regime.}
\label{fig6}
\end{figure}

We have also tried two different evolution histories. The first one is in the strong EM coupling regime with $n$ changing from $n=-0.1$ to $n=-4.7$ at $N(t)=31.5$. The constant parameter is $p=0.1$ and the parameter $c$ is time dependent in such a way that $c=1$ for $N(t)=3.15-10$ and otherwise $c=0$. The second one is in the weak EM coupling regime with $n$ changing from $n=0$ to $n=10$ at $N(t)=39$, $p=0$, and $c$ changing from $c=0$ to $c=500$ at $N(t)=3$. The time evolutions of the electromagnetic energy densities for the first and the second cases are shown in Fig.~\ref{fig3} and Fig.~\ref{fig5}, respectively. The spectra of the generated PMF for both cases are shown in Fig.~\ref{fig4} and Fig.~\ref{fig6}, respectively. Once again, we have seen how the $J(\tau)$ term  allows the modes of comologically interesting scales to grow sufficiently fast at a much later time. However, in the weak EM coupling regime, the generated PMF is still too small to be the seed fields.

Many papers have studied the problem of magnetogenesis during inflation spoiling the cosmological density perturbations~\cite{constraint1, constraint2} and generating primordial tensor perturbations (Ref.~\cite{sorbo} and references therein). The studies most relevant to the present consideration are the constraints from measurements of the power spectrum and bispectrum of the cosmic microwave background~\cite{constraint2}, in which the authors have considered the above mentioned model $I(\tau)=1$ and $J(\tau)=\alpha\phi/f$ as shown in Eq.~(\ref{axionXi}) and derived an upper limit on a combined parameter: $\xi<2.4$. This limit translates into the condition that during inflation $\rho_{EM}/(\rho_\phi+\rho_\gamma) < 10^{-8}$. In Fig.~\ref{fig1} and Fig.~\ref{fig3}, where we have successful magnetogenesis, $\rho_{EM}/(\rho_\phi+\rho_\gamma)$ is much less than $10^{-8}$ for the time when EM fields are generated during inflation. When the modes of comologically interesting scales grow at $N\sim 30-50$, $\rho_{EM}/(\rho_\phi+\rho_\gamma) \sim 10^{-13}$ in Fig.~\ref{fig1} and $\rho_{EM}/(\rho_\phi+\rho_\gamma) \sim 10^{-14}$ in Fig.~\ref{fig3}. It is worth to study in details the effects of magnetogenesis in the present work on the cosmic microwave background.

\section{Conclusion}

It is amazing that the Universe has been so good a conductor that a seed magnetic field can hardly be generated in it. But once seed magnetic fields were generated by a whatsoever mechanism in the early Universe, they were kept frozen and preserved in the conducting plasma state of the Universe. There has been a lot of proposals for generating magnetic fields in the cosmic plasma via astrophysical magnetogenesis. However, it was thought that the best way to avoid the high conductance is to generate primordial magnetic fields from photon quantum fluctuations during inflation when the Universe was in a vacuum state. Unfortunately, this thought failed due to the conformal invariance of electromagnetism. 

So far, most common models for inflationary magnetogenesis involve the breaking of conformal invariance in the context of dilaton electromagnetism. However, a successful model would require a strong electromagnetic coupling regime that cannot be trusted physically. These models have been extended by including an axion-photon coupling. This coupling can generate and add helicity to a blue spectrum of magnetic fields, so they still rely on the inverse cascade to convert shorter-wavelength magnetic fields to cosmologically interesting ones.

In the present work, we have pointed out that the main obstacle in inflationary magnetogenesis is the weakening of the generated large-scale magnetic fields by a hugh conformal factor due to the inflationary dilution. In light of this, we have generalized the axion dynamics to a time dependent coupling $J(a) F\tilde{F}/4$. We have found that in order to generate large-scale magnetic fields during inflation, the $J(a)$ function must be able to delay their growth to a later stage of inflation. Thus we have given a recipe of the forms of the $I(a)$ and $J(a)$ functions for which a successful inflationary magnetogenesis can be made, though we admit that it is rather contrived. It seems that the axion dynamics may involve a specific form of the axion potential or even the evolution of extra dimensions in string-inspired theories. Then it would be interesting to build a particle and field model to realize the idea of using $I$-$J$ functions proposed in this work. Furthermore, a complete analysis including the backreaction and observational constraints is in order.

\begin{acknowledgments}
This work was supported in part by the National
Science Council, Taiwan, ROC under the Grant No.
NSC101-2112-M-001-010-MY3 (K.W.N.) and the Office of Research and Development,
National Taiwan Normal University, Taiwan, ROC
under the Grant No. 102A05 (W.L.).
\end{acknowledgments}

\end{document}